\documentclass[aps,superscriptaddress,nofootinbib,eqsecnum,prd,notitlepage,twocolumn,showkeys]{revtex4-1} 

\pdfoutput=1

\usepackage{amsfonts}
\usepackage{amsmath}
\usepackage{amssymb}
\usepackage{graphicx,color}
\usepackage{float}
\usepackage{hyperref}
\usepackage{subfigure}
\usepackage{dcolumn}
\usepackage{soul}
\usepackage{ulem}
\usepackage{verbatim}

\begin{document}

\title{Embedding Ultra slow-roll inflaton dynamics in Warm Inflation}
\author{Sandip Biswas}
\email{sandipb20@iitk.ac.in}
\affiliation{Department of Physics, Indian Institute of Technology, Kanpur: 208016,
Uttar Pradesh, India}

\author{Kaushik Bhattacharya}
\email{kaushikb@iitk.ac.in}
\affiliation{Department of Physics, Indian Institute of Technology, Kanpur: 208016,
Uttar Pradesh, India}

\author{Suratna Das}
\email{suratna.das@ashoka.edu.in}
\affiliation{Department of Physics, Ashoka University,
   Rajiv Gandhi Education City, Rai, Sonipat: 131029, Haryana, India}

\begin{abstract}

Slow-roll of the inflaton field defines the standard dynamics of the inflationary epoch. However, the inflationary system deviates from slow-roll when it encounters an extremely flat region of the inflaton potential, and enters a phase dubbed Ultra slow roll. In this article, we explore the possibility of realizing an Ultra slow-roll phase in a particularly interesting inflationary scenario, called Warm Inflation. In the Warm inflationary scenario a thermalized, sub-dominant radiation bath coexists with the inflaton energy density as an effect of dissipative dynamics. We show in this article that though the background dynamics indicate Ultra slow-roll when the potential becomes extremely flat, in Warm Inflation models, where the dissipation coefficient is a sole function of the temperature of the radiation bath, the system fails to maintain the thermal equilibrium as soon as it enters the Ultra slow-roll phase. As thermal equilibrium is a key feature of Warm Inflation, and as it is not yet known how to deal with Warm Inflation without thermal equilibrium, we could not analyze such systems any further in this article. However, we demonstrate that brief periods of Ultra slow-roll phase, which smoothly ends into standard slow-roll, can be accommodated in WI models where the dissipation coefficient is not only a function of the temperature of the radiation bath but also depends on the amplitude of the inflaton field. We theoretically determine the criteria of successfully embedding Ultra slow-roll in WI while the system remain in thermal equilibrium, and also demonstrate numerically that such short Ultra slow-roll phases can indeed be embedded in specific Warm Inflation models which comply with the theoretically determined criteria. 

\end{abstract}

\maketitle

\section{Introduction}

Cosmic Inflation \cite{Kazanas:1980tx, Guth:1980zm, Sato:1981ds, Sato:1980yn, Linde:1981mu, Albrecht:1982wi}, a brief period of near-exponential expansion of the early universe, has now become an integrated part of the standard big bang cosmology that not only solves the fine-tuning problems of the hot big bang model but also helps generating seeds for the large-scale structures we see today. In the standard scenario, cosmic inflation is driven by the potential energy of a slow-rolling scalar field, dubbed inflaton. To maintain the slow-rolling of the field during inflation, the potential of the inflaton field needs to be sufficiently flat. However, there are regimes when the slow-rolling conditions cannot remain valid. One such situation occurs when the potential becomes extremely flat, such as around inflection points. The system then deviates from the standard slow-rolling and enters a phase of so-called ``Ultra slow-roll'' \cite{Dimopoulos:2017ged}. The term `Ultra slow-roll' was first coined in \cite{Kinney:2005vj}, and the dynamics of the system during such a phase were studied. However, a similar situation was first investigated in \cite{Inoue:2001zt}. The attractive behaviour and stability of an Ultra slow-roll phase were investigated in \cite{Pattison:2018bct}. Such a departure from slow-roll often turns out to be useful as such a phase allows the cosmological perturbations to grow sufficiently enough to generate Primordial Black Holes \cite{Motohashi:2017kbs}, a viable candidate for Dark Matter \cite{Carr:2016drx}.\footnote{An issue regarding whether or not the large-scale cosmological perturbations, probed by the Cosmic Microwave Background, receive large non-perturbative corrections from the enhanced perturbations on small scales due to Ultra slow-roll is being debated in the literature \cite{Kristiano:2022maq, Riotto:2023hoz, Kristiano:2023scm, Riotto:2023gpm, Franciolini:2023lgy, Choudhury:2023vuj}. If they do, then such mechanisms of generating Primordial Black Holes will be ruled out. The issue is yet to be resolved.} 

In the standard inflationary scenario, the couplings of the inflaton field with other particles are considered to be negligible. Thus, all other energy densities present before inflation are diluted away exponentially by the time inflation ends, and, therefore, to onset the standard hot big bang evolution post-inflation, a separate phase of reheating \cite{Kofman:1994rk} is called for. This standard inflationary scenario will be referred to as `Cold Inflation' (CI) in this article. However, there is an alternate inflationary scenario, dubbed `Warm Inflation' (WI) \cite{Berera:1995ie} (for recent reviews on WI, see, e.g., \cite{Kamali:2023lzq, Berera:2023liv}), where those couplings of the inflaton field to other particles play a significant role during the inflationary evolution in dissipating inflaton's energy density into a subdominant, yet non-negligible, radiation bath. As a constant radiation bath is maintained throughout WI, this helps WI to transit smoothly into a radiation dominated universe once WI ends. Thus, WI, unlike CI, does not call for a reheating phase post inflation. Moreover, despite being sub-dominant, the produced radiation energy density, $\rho_r$, satisfies the condition $\rho_r^{1/4}>H$ (where $H$ is the Hubble parameter during inflation), which upon assuming thermalization of the radiation bath yields the condition $T>H$ ($T$ being the temperature of the thermalized radiation bath). This thermalization condition is maintained throughout the evolution of WI, and plays a major role in determining the cosmological perturbations produced during WI \cite{Kamali:2023lzq, Berera:2023liv}. Graceful exit of WI, which is a much more complex process than in CI, has been extensively studied in \cite{Das:2020lut}. 

WI has certain attractive features than its more conventional counterpart, CI. First of all, as mentioned above, WI is not in need of an extra reheating phase --- physics of which is not yet fully understood. Secondly, WI yields a more enhanced scalar curvature power spectrum compared to CI \cite{Hall:2003zp, Ramos:2013nsa}, which lowers the tensor-to-scalar ratio significantly. This allows WI to accommodate potentials, such as quartic self-coupling potentials ($\lambda\phi^4$), which are otherwise ruled out in CI for generating way too much tensor-to-scalar ratios \cite{Bartrum:2013fia}. It has been shown in \cite{Bastero-Gil:2011clw} that the observed baryon asymmetry in the nature can be explained by the dissipative effects of the WI alone, which is absent in CI. This also leads to observable baryon isocurvature perturbations which can help in checking the consistency of the WI models \cite{Bastero-Gil:2014oga}. Certain WI models can also generate Primordial Black Holes without invoking any departure from slow-roll \cite{Arya:2019wck, Bastero-Gil:2021fac, Correa:2022ngq, Arya:2023pod}. Moreover, it has been shown recently that, while CI fails to comply \cite{Kinney:2018nny} with the de Sitter Swampland Conjecture in String Theory proposed in \cite{Ooguri:2018wrx, Garg:2018reu}, one can easily overcome the obstacles in WI due to its very construction and can successfully accommodate the criteria of the conjecture within the framework of WI \cite{Das:2018hqy, Motaharfar:2018zyb, Das:2018rpg, Das:2019hto, Das:2019acf, Das:2020xmh}. Therefore, WI is preferred over CI as an inflationary paradigm in low energy effective field theories which decent from ultraviolet complete theories of gravity, such as String Theory.

In this article, we investigate what happens to the dynamics of WI when it encounters an extremely flat region of a potential, like an inflection point. We discussed before that, in CI, the system significantly deviates from slow-rolling and enters a phase of Ultra slow-roll in a similar situation. Thus, can we expect a similar behaviour in WI as well? We found that though the background dynamics does show signatures of Ultra slow-roll, the system deviates from its thermal equilibrium exponentially fast in models of WI where the dissipative coefficient is a sole function of the temperature of the radiation bath. As thermal equilibrium of the system is a key feature of WI, it is not clear what happens to the dynamics of WI if the thermal equilibrium is lost during any phase of its evolution. Therefore, we could not analyze such systems any further in this article. However, we showed that WI models with dissipative coefficients depending on the temperature of the thermal bath as well as on the amplitude of the inflaton field can successfully realize brief periods of Ultra slow-roll while maintaining the overall thermalization of the system. 

We have organized the rest of the article as follows. In Sec.~(\ref{CI-USR}), we analyze the Ultra slow-roll dynamics in standard CI and define the criteria which distinguish Ultra slow-roll from slow-roll in terms of Hubble slow-roll parameters. In Sec.~(\ref{WI-SR}), we briefly discuss the WI dynamics under slow-roll and show how the thermalization of the system is maintained during the slow-roll phase. In Sec.~(\ref{WI-USR}) we determine the theoretical criteria which can lead to an Ultra slow-roll phase in WI while maintaining thermal equilibrium of the system. This follows by the Sec.~(\ref{numerical}), where we analyze numerically a specific WI model (which can accommodate an Ultra slow-roll phase according to the theoretical criteria developed in Sec.~(\ref{WI-USR})) with two potentials with extremely flat regions, and show that brief periods of Ultra slow-roll phases can indeed be realized while maintaining the thermalization of the system. We also show that the system smoothly enters a standard slow-roll phase after these brief periods of Ultra slow-roll. In Sec.~(\ref{conclusion}), we discuess the main results obtained in this article and then conclude.

\section{Ultra slow-roll in Cold Inflation}
\label{CI-USR}

To figure out how WI behaves when the potential becomes extremely flat, we need to first understand how the dynamics deviates from slow-roll (and enters an Ultra slow-roll phase) in CI in a similar situation. In this section we will closely follow the arguments given in \cite{Dimopoulos:2017ged}, and, in the following section, we will generalise the arguments presented here for the case of WI. 

In canonical CI models, a single scalar inflaton field, $\phi$, evolves according to the Klein-Gordon equation given as 
\begin{eqnarray}
\ddot\phi+3H\dot\phi+V,_\phi=0,
\label{KGeq}
\end{eqnarray}
where the overdot denotes derivative with respect to the cosmic time $t$ and $V,_\phi=dV/d\phi$. 
Following \cite{Dimopoulos:2017ged}, we will call the three terms in the above equation as the acceleration term, the friction term, and the slope term, respectively. The Friedmann-Lema\^{i}tre-Robertson-Walker scale factor, $a(t)$, evolves according to the Friedmann equations:
\begin{eqnarray}
3M_{\rm Pl}^2H^2&=&\frac{\dot\phi^2}{2}+V(\phi), \label{friedmann-1}\\
2M_{\rm Pl}^2\dot H&=&-\dot\phi^2,
\label{friedmann-2}
\end{eqnarray}
where $H\equiv \dot{a}/a$ is the Hubble parameter and $M_{\rm Pl}$ is the reduced Planck mass. The background evolution is characterized in terms of the Hubble slow-roll parameters defined by \cite{Martin:2012pe}:
\begin{eqnarray}
\epsilon_{i+1}=\frac{d\ln \epsilon_i}{dN},
\end{eqnarray}
where $N\equiv \ln a$ denotes the number of e-folds. Starting with $\epsilon_0\propto1/H$, the consecutive first slow-roll parameter $\epsilon_1$ and the second slow-roll parameter $\epsilon_2$ can be expressed as 
\begin{eqnarray}
\epsilon_1&\equiv&-\frac{\dot H}{H^2}, \label{epsilon1}\\
\epsilon_2&\equiv&\frac{\dot\epsilon_1}{\epsilon_1H}=2\epsilon_1+\frac{\ddot H}{H\dot H}. 
\label{epsilon2}
\end{eqnarray}
Inflation requires $\epsilon_1<1$, and the validity of slow-roll approximation is ensured by the general conditions $|\epsilon_i|\ll1$. During inflation the potential term dominates over the kinetic term. Thus, from the Friedmann equations given in Eq.~(\ref{friedmann-1}) and Eq.~(\ref{friedmann-2}, we see that $\epsilon_1\sim 3\dot\phi^2/(2V(\phi))$, ensuring $\epsilon_1\ll1$ during inflation. And during slow-roll, the acceleration term ($\ddot\phi$) is negligible compared to the friction ($H\dot\phi$) and the slope ($V,_\phi$) terms in the Klein-Gordan equation given in Eq.~(\ref{KGeq}), which then can be approximated as 
\begin{eqnarray}
3H\dot\phi+V,_\phi\simeq 0. 
\end{eqnarray}
This suggests that 
\begin{eqnarray}
\frac{\ddot H}{H\dot H}=\frac{2\ddot\phi}{H\dot\phi}\simeq2\epsilon_1-2\eta_V,
\end{eqnarray}
where $\eta_V$ is one of the two potential slow-roll parameters: 
\begin{eqnarray}
\epsilon_V\equiv \frac{M_{\rm Pl}^2}{2}\left(\frac{V,_\phi}{V}\right)^2, \quad\quad \eta_V \equiv M_{\rm Pl}^2\frac{V,_{\phi\phi}}{V}.
\label{pot-slow-roll}
\end{eqnarray}
Both these potential slow-roll parameters quantify the flatness of the inflaton potential during inflation. As the potential requires to be nearly flat for the slow-rolling of the field, the potential slow-roll parameters require to be much smaller than unity during slow-roll inflation. Therefore, from Eq.~(\ref{epsilon2}) we see that $|\epsilon_2|\sim |4\epsilon_1-2\eta_V|\ll1$. Thus, both $\epsilon_1$ and $|\epsilon_2|$, being much smaller than unity, ensure the slow-roll dynamics of the inflaton field. 

However, when the potential becomes extremely flat, the slope term in the Klein-Gordan equation becomes negligible, and it becomes 
\begin{eqnarray}
\ddot\phi+3H\dot\phi\simeq0,
\end{eqnarray}
yielding $\epsilon_2\sim -6+2\epsilon_1$. We note that $\epsilon_1$ remains much smaller than unity even when the potential becomes extremely flat, as the potential term dominates over the kinetic term. Therefore, inflation does not stop when the potential becomes extremely flat. However, $\epsilon_2$ becomes of the order of unity ($\epsilon_2\sim-6$). This clearly indicates that the scalar field dynamics deviates from slow-roll when the potential becomes extremely flat, and enters a new phase of evolution, dubbed the Ultra slow-roll. 


\section{Warm Inflation: The slow-roll regime}
\label{WI-SR}

During WI, the inflaton field, $\phi$, dissipates its energy to a radiation bath, maintaining a non-negligible radiation energy density, $\rho_r$, throughout. This feature distinguishes WI from the standard CI scenario. Therefore, the equation of motion of the inflaton field also differs from that of in the CI scenario. The equations governing the dynamics of the inflaton field, $\phi$, and the radiation bath, $\rho_r$, in WI can be written as 
\begin{eqnarray}
&&\ddot\phi+3H\dot\phi+V,_\phi=-\Upsilon(\phi,T)\dot\phi,
\label{KG-WI}\\
&&\dot\rho_r+4H\rho_r=\Upsilon(\phi,T)\dot\phi^2.
\label{rad-bath-eq}
\end{eqnarray}
Here, $\Upsilon$ is the dissipative term which can depend on the amplitude of the inflaton field, $\phi$, as well as the temperature of the radiation bath, $T$. It is assumed that the radiation bath, generated by the dissipation of the inflaton field, is in near thermal equilibrium throughout WI, and thus a temperature $T$ can be defined. Many microphysical models of Warm Inflation have been studied over the years and the decay rates (under the assumption of thermal equilibrium) have been calculated. For such studies one can look into \cite{{Bastero-Gil:2010dgy, Berera:2008ar}} and \cite{Kamali:2023lzq} and for a more recent review. Form these studies, one can see that a general form of the dissipative coefficient in Warm Inflation can be written in the form
\begin{eqnarray}
\Upsilon(\phi,T)=C_\Upsilon T^p\phi^c M^{1-p-c},
\label{Ups}
\end{eqnarray}
where $C_\Upsilon$ is a dimensionless constant carrying the signatures of the microscopic model used to derive the dissipative coefficient (such as the different coupling constants), and $M$ is some appropriate mass-scale, so that the dimensionality of the dissipative coefficient is preserved, $[\Upsilon]=$[mass] in our system of units where $\hbar=c=1$. The numerical powers of $T$ and $\phi$, which are $p$ and $c$ respectively, can take positive or negative values, however, we will restrict ourselves to $|p|<4$ due to the stability of the WI models \cite{Moss:2008yb, Bastero-Gil:2012vuu}.

We define the dimensionless parameter $Q$ as
\begin{eqnarray}
Q\equiv \frac{\Upsilon}{3H},
\end{eqnarray}
which is the ratio of the two friction terms appearing in the equation of motion of the inflation field, one due to dissipation ($\Upsilon\dot\phi$) and the other due to the Hubble expansion ($3H\dot\phi$). If the Hubble friction dominates over the friction due to dissipation, i.e. $Q<1$, we call it a weak dissipative regime of WI. On the other hand, when the friction due to dissipation dominates the equation of motion of the inflaton field ($Q>1$ in such cases), we call it a strong dissipative regime of WI. 

Apart from the two potential slow-roll parameters defined in Eq.~(\ref{pot-slow-roll}), there are two additional slow-roll parameters in WI \cite{Hall:2003zp, Berera:2008ar}:
\begin{eqnarray}
\beta\equiv M_P^2\frac{\Upsilon_{,\phi}V_{,\phi}}{\Upsilon V}, \quad\quad \delta\equiv\frac{TV_{,\phi T}}{V_{,\phi}},
\end{eqnarray}
which are required to ensure the slow-roll dynamics. During slow-roll $\epsilon_V$, $\eta_V$, $\beta$ and $\delta$ are all smaller than $1+Q$ in both weak and strong dissipative regime. 

In the slow-roll regime, i.e. when the acceleration term in Eq.~(\ref{KG-WI}) is negligible with respect to all the other terms, we can approximate this equation as
\begin{eqnarray}
3H(1+Q)\dot\phi+V,_\phi\approx0.
\end{eqnarray}
However, as the inflaton potential energy density always dominates over the kinetic term and the radiation energy density during WI, the Friedmann equation given in Eq.~(\ref{friedmann-1}) can be written as 
\begin{eqnarray}
3M_{\rm Pl}^2H^2\approx V(\phi).
\label{friedmann-approx}
\end{eqnarray}
As during WI a constant radiation bath is maintained by the dissipation of energy of the inflaton field into the radiation bath, we can assume $\dot\rho_r\approx0$, and can approximate Eq.~(\ref{rad-bath-eq}) as 
\begin{eqnarray}
4H\rho_r\approx \Upsilon\dot\phi^2.
\label{const-rad-bath}
\end{eqnarray}
Also, as the radiation bath is near equilibrium, we can define a temperature $T$ of the radiation bath as
\begin{eqnarray}
\rho_r=\frac{\pi^2}{30}g_*T^4,
\label{rad-T}
\end{eqnarray}
where $g_*$ is the relativistic degrees of freedom of the radiation bath. We can then determine the evolution of $T$ with respect to the e-foldings $N$ during slow-roll as 
\begin{eqnarray}
\frac{d\ln T}{dN}=\frac{\dot T}{HT}\simeq\frac{1}{4-p}\left(\frac{\epsilon_V}{1+Q}-\frac{\beta}{1+Q}\right)\ll1.\nonumber\\
\label{T-rate-SR}
\end{eqnarray}  
Here we have used the relation $dN=Hdt$. 
The above equation ensures that during slow-roll, the temperature of the radiation bath evolves very slowly, maintaining the near equllibrium condition. 

With all these conditions of slow-roll WI, we find that during a slow-roll phase in WI 
\begin{eqnarray}
\epsilon_1 &=& \frac{\epsilon_V}{1+Q},\\
\epsilon_2 &=& -2\frac{\eta_V}{1+Q}+ \left(\frac{4+3Q}{1+Q}-\frac{p}{4-p}\frac{Q}{1+Q}\right)\frac{\epsilon_V}{1+Q} \nonumber\\
&&+\left(\frac{4}{4-p} \frac{Q}{1+Q}\right)\frac{\beta}{1+Q}.
\label{eps2wrsr}
\end{eqnarray}
In deriving these relations we have used the general form of the dissipative coefficient given in Eq.~(\ref{Ups}).
As $\epsilon_V$, $\eta_V$ and $\beta$ are all smaller than $1+Q$ during slow-roll we see that $\epsilon_1$ and $|\epsilon_2|$ are both smaller than unity during slow-roll. In the next section we will see that this situation will change when the potential becomes extremely flat, as it happens in CI.

\section{Realising Ultra slow-roll within Warm Inflation}
\label{WI-USR}

We note in Sec.~(\ref{CI-USR}) that in CI, the evolution enters an Ultra slow-roll phase when the potential becomes extremely flat, and we quantify it by showing that the magnitude of the second Hubble slow-roll parameter, $\epsilon_2$, becomes of the order of unity, while the first Hubble slow-roll parameter, $\epsilon_1$, remains much smaller than 1. Does a similar situation arise in WI when the potential becomes extremely flat? While answering this question, one also needs to keep in mind that WI maintains a radiation bath in near equilibrium throughout the slow-roll evolution. Can WI enter an Ultra slow-roll phase while maintaining the thermal equilibrium of the radiation bath present during the evolution? 

To answer these questions, we first determine the evolution of the temperature $T$ for a phase when the potential becomes extremely flat. During such an evolution, as $V,_\phi\approx0$, we can approximate Eq.(\ref{KG-WI}) as 
\begin{eqnarray}
\ddot\phi+3H(1+Q)\dot\phi\approx0.
\label{KG-approx}
\end{eqnarray}
It is to note that an Ultra slow-roll phase during an inflationary epoch, both in CI and WI, must be a transient phase, i.e. the Ultra slow-roll phase must be preceded by a slow-roll phase and ends into another slow-rolling phase. Thus the initial conditions of the Ultra slow-roll phase is set by the preceding slow-roll phase. In WI, a thermal equilibrium is assumed in the slow-rolling phase preceding the Ultra slow-roll phase. Therefore, the thermalization condition and the form of the dissipative coefficient during Ultra slow-roll phase are set by the preceding slow-roll phase. With these initial conditions, we can then let the system evolve during an Ultra slow-roll phase during WI.  Hence, we will assume that at the onset of such an evolution a constant radiation bath is maintained, and that radiation bath will remain near thermal equilibrium so that a temperature $T$ can be defined. In such a case, both Eq.~(\ref{const-rad-bath}) and Eq.~(\ref{rad-T}) would remain valid. From Eq.(\ref{KG-approx}), Eq.~(\ref{const-rad-bath}) and Eq.~(\ref{rad-T}) it is straightforward to show that the temparature $T$ evolves as
\begin{eqnarray}
\frac{d\ln T}{dN}=\frac{1}{4-p}\left[c\frac{\dot\phi}{H\phi}+\epsilon_1-6(1+Q)\right]\,,
\label{T-rate-USR}
\end{eqnarray}
when the potential becomes extremely flat. We can see here that the first and the last terms on the right hand side of the above eqution are not proportional to any slow-roll parameters and can take large values. Thus, comparing this equation with Eq.~(\ref{T-rate-SR}), which depicts the evolution of $T$ during slow-roll in WI, we see that one cannot conclude right away that the temperature will evolve slowly during the evolution through an extremely flat region of the potential. Can we make the temperature evolve slowly in such a case so that the system can evolve being near equilibrium? We will address this question by the end of this section. 

We will now see what happens to the first Hubble slow-roll parameter, $\epsilon_1$, when the potential becomes extremely flat. We note that, in WI, 
$-2M_{\rm Pl}^2\dot H=\rho_\phi+P_{\phi}+\rho_r+P_r=\dot\phi^2+(3/4)\rho_r,$
where $\rho_\phi$ and $P_\phi$ are the energy density and the pressure of the inflaton field, and similarly, $\rho_r$ and $P_r$ are the radiation energy density and radiation pressure, respectively. However, using Eq.~(\ref{const-rad-bath}), we can write 
\begin{eqnarray}
-2M_{\rm Pl}^2\dot H=(1+Q)\dot\phi^2.
\label{Hdot-WI}
\end{eqnarray}
Thus, using the above equation and Eq.~(\ref{friedmann-approx}), the first Hubble slow-roll parameter can be expressed as 
\begin{eqnarray}
\epsilon_1\sim \frac32 \frac{(1+Q)\dot\phi^2}{V(\phi)}.
\end{eqnarray}
Now, as during inflation the potential energy density dominates over kinetic energy density, we can see that in the weak dissipative regime ($Q<1$) $\epsilon_1$ can remain much smaller than unity ensuring inflation to continue even when the potential becomes extremely flat. However, in strong dissipation  regime, there are models where the $Q$ values can be as large as $\mathcal{O}(10^3)$ \cite{Berghaus:2019whh, Das:2019acf, Das:2020xmh}. In such models, it might happen that when the potential becomes extremely flat, $Q\dot\phi^2\sim V(\phi)$, yielding $\epsilon_1\sim1$, which is an indication of end of inflation. But, there are WI models realised in strong dissipation \cite{Bastero-Gil:2019gao} where $Q$ is of the order of 10 or 100. In such models, inflation continues to take place even when the potential becomes extremely flat. 

Now, to determine the second Hubble parameter, $\epsilon_2$, during such an evolution, we use Eq.~(\ref{KG-approx}), Eq.~(\ref{T-rate-USR}) and Eq.~(\ref{Hdot-WI}) to show that 
\begin{eqnarray}
\frac{\ddot{H}}{\dot{H}H}&=&-6(1+Q)+\frac{4}{4-p}\frac{\Upsilon_{,\phi}\dot{\phi}}{3H^{2}(1+Q)}
\nonumber\\
&&+\left[ \frac{1}{4-p}\frac{\Upsilon_{,T} T}{\Upsilon} + 1\right]\frac{Q}{1+Q}\epsilon_1 -\dfrac{6Qp}{4-p}\,.
\end{eqnarray}
Using this relation into Eq.~(\ref{epsilon2}), we see that 
\begin{eqnarray}
\epsilon_2
=-6\left(1+\frac{4Q}{4-p}\right)+\frac{4}{4-p}\left[c\frac{\dot\phi}{H\phi}+\epsilon_1\right]\frac{Q}{1+Q}+2\epsilon_1,\nonumber\\
\label{eps2usr}
\end{eqnarray}
where we have again used the general form of the dissipative coefficient given in Eq.~(\ref{Ups}). Let us appraise the situation in two different regimes, strong dissipation and weak dissipation, separately. 

In strong dissipative regime ($Q\gg1$), Eq.~(\ref{eps2usr}) can be approximated as 
\begin{eqnarray}
\epsilon_2\sim\frac{4}{4-p}\left[-6Q+c\frac{\dot\phi}{H\phi}+\epsilon_1\right]+2\epsilon_1\,.
\label{eps2-strong}
\end{eqnarray}
There are couple of models where WI can be realised in strong dissipative regime: one is presented in  \cite{Bastero-Gil:2019gao}, and the other is presented in \cite{Berghaus:2019whh}  (this model is dubbed Minimal Warm Inflation). In the strong dissipative WI model presented in  \cite{Bastero-Gil:2019gao}, the dissipative coefficient, despite having a complex form, varies inversely with the temperature ($\Upsilon\propto T^{-1}$) during the evolution, and does not depend on the inflaton amplitude. Therefore, for this kind of model $p=-1$ and $c=0$. This yields $\epsilon_2\sim (-24/5)Q+(14/5)\epsilon_1$. We have mentioned earlier that in this model, $Q$ can be of the order of 10 or 100 (yielding $\epsilon_1\ll1$), and thus the model can enter an Ultra slow-roll phase while inflating. However, looking at Eq.~(\ref{T-rate-USR}), we note that $d\ln T/dN\sim -(6/5)Q$, and therefore, the temperature decreases exponentially with the e-foldings during the Ultra slow-roll phase. It indicates that the thermal equilibrium of the radiation bath cannot be maintained during Ultra slow-roll in such models, and the system will deviate from the basic WI picture. In the Minimal Warm Inflation model \cite{Berghaus:2019whh}, the dissipative coefficient varies with the cubic power of temperature ($\Upsilon\propto T^3$) and does not depend on the inflaton amplitude. Therefore in this model $p=3$ and $c=0$, yielding $\epsilon_2\sim -24Q+6\epsilon_1$. As mentioned earlier, $Q$ can take very large values ($\mathcal O (10^3)$) in such models. Despite that, if $Q\phi^2\ll V(\phi)$, then $\epsilon_1\ll1$ and we get $\epsilon_2\sim-24 Q$, indicating an Ultra slow-roll phase. Even though, like in the previous case, we note that $d\ln T/dN\sim-6Q$, and the temperature will exponentially fall with e-foldings taking the system away from thermal equilibrium, and deviating from the standard WI picture. 

In weak dissipative regime $(Q\ll1)$, Eq.~(\ref{eps2usr}) can be approximated as 
\begin{eqnarray}
\epsilon_2=-6+\frac{4}{4-p}\left[c\frac{\dot\phi}{H\phi}+\epsilon_1\right]Q+2\epsilon_1\,.
\label{eps2-weak}
\end{eqnarray}
We will deal with two kinds of models where WI is realised in weak dissipative regime. The first one is dubbed Warm Little Inflaton \cite{Bastero-Gil:2016qru}, where the dissipative coefficient varies linearly with the temperature ($\Upsilon\propto T$) and does not depend on the inflaton amplitude. Therefore, in this case, $p=1$ and $c=0$, yielding $\epsilon_2=-6+(10/3)\epsilon_1$. As we have seen before, $\epsilon_1\ll1$ when the potential becomes extremely flat during weak dissipation, and therefore, we conclude that during such a phase $\epsilon_2\sim-6$. Though it indicates an onset of an Ultra slow-roll phase, we again note that, as in the strong dissipative cases discussed above, Eq.~(\ref{T-rate-USR}) yields $d\ln T/dN\sim-2$, indicating departure from thermal equilibrium. So far, we have discussed WI models where the dissipative coefficient varies solely with $T$, both in strong and weak dissipative regimes, and saw that the system veers away from the thermal equilibrium state when an Ultra slow-roll like phase sets in. As it is not yet known how to treat WI when the thermal equilibrium of the radiation bath is lost, we will not further analyze the Ultra slow-roll phase in WI models where the dissipative coefficient solely varies with the temperature and has no dependence on the field amplitude. 

We will now analyze a WI model where the dissipative coefficient is a function of both $T$ and $\phi$. 
The model studied in \cite{Berera:2008ar, Bastero-Gil:2010dgy, Bastero-Gil:2012akf}  has a dissipative coefficient of the form $\Upsilon\propto T^3/\phi^2$, and WI is realised in weak dissipative regime in such a model. This model has also been verified with the Planck data in \cite{Benetti:2016jhf, Arya:2017zlb}. Being realized in weak dissipative regime, we can guarantee that $\epsilon_1$ will remain much smaller than unity even when the potential becomes extremely flat. Our aim, now, is to keep $|d\ln T/dN|\approx 0$ while making $|\epsilon_2|$ larger than unity, so that an Ultra slow-roll phase can be realized while maintaining the thermal equilibrium of the system. Therefore, with $\epsilon_1\ll1$ and $Q\ll1$, Eq.~(\ref{T-rate-USR}) can be approximated as 
\begin{eqnarray}
\frac{d\ln T}{dN}\approx \frac{1}{4-p}\left(c\frac{d\ln\phi}{dN}-6\right). 
\end{eqnarray}
We note from this equation that, to keep $|d\ln T/dN|\approx 0$, $d\ln\phi/dN$ should be positive if $c$ is positive and vice versa. In other words, $|c||d\ln\phi/dN|$ should always remain positive. Hence, to maintain a thermal equilibrium we demand that during an Ultra slow-roll phase 
\begin{eqnarray}
\left|\frac{d\ln\phi}{dN}\right|\sim\frac{6}{|c|}. 
\label{thermal-condition}
\end{eqnarray}
If we impose this condition on $\epsilon_2$ given in Eq.~(\ref{eps2-weak}) with other conditions, like $\epsilon_1\ll1$ and $Q\ll1$, we obtain 
\begin{eqnarray}
\epsilon_2\approx-6+\left(\frac{4}{4-p}\right)6Q.
\end{eqnarray}
We can keep $|\epsilon_2|>1$ in two ways: keeping $\epsilon_2>1$ or demanding $\epsilon_2<-1$. 
In the first case, when $\epsilon_2>1$, we get 
\begin{eqnarray}
Q>\frac76\left(\frac{4-p}{4}\right).
\end{eqnarray}
Therefore, to ensure that WI takes place in weak dissipative regime we constrain $Q$ as $(7/6)[(4-p)/4]<Q<1$. For the model discussed above \cite{Berera:2008ar, Bastero-Gil:2010dgy, Bastero-Gil:2012akf},  this condition leads to $(7/24)<Q<1$. However, in the second case, when  $\epsilon_2<-1$, we get 
\begin{eqnarray}
Q<\frac56\left(\frac{4-p}{4}\right).
\end{eqnarray}
This condition ensures that WI will take place in weak dissipative regime. For the model discussed above \cite{Berera:2008ar, Bastero-Gil:2010dgy, Bastero-Gil:2012akf},  this condition leads to $Q<5/24$.

However, we note that though the condition given in Eq.~(\ref{thermal-condition}) allows the onset of Ultra slow-roll in WI, while maintaining thermal equilibrium,  still Eq.~(\ref{thermal-condition}) differs from the equation of motion of the inflaton field during Ultra slow-roll given in Eq.~(\ref{KG-approx}). Therefore, as Ultra slow-roll proceeds, the dynamics of the inflaton field will take the system away from the condition in Eq.~(\ref{thermal-condition}), and the temperature of the system will start to evolve indicating a departure from thermal equilibrium. Thus, the system needs to exit from Ultra slow-roll before the temperature evolves too much to disrupt thermal equilibrium. We will show in the next section, by numerically evolving the system, that such a thermally equilibrated Ultra slow-roll phase can be realised in WI in specific cases. 

\section{Numerical analysis of viable Ultra slow-roll phase in WI}
\label{numerical}

In WI, the inflaton field equation of motion and the evolution of the radiation bath (as well as the evolution of the temperature) are coupled as the inflaton field dissipates its energy to the radiation bath throughout WI. The system thus evolves according to the coupled equations given in Eq.~(\ref{KG-WI}) and Eq.~(\ref{rad-bath-eq}). We will numerically evolve these two equations in cases where WI can enter an Ultra slow-roll phase, and then appraise its characteristics. We will consider two different potentials, the linear potential \cite{Dimopoulos:2017ged}, and the cubic potential \cite{Pattison:2018bct}, where it has been shown previously that the system undergoes Ultra slow-roll in CI scenarios. In both these cases, we will make use of the dissipative coefficient $\Upsilon=C_\Upsilon T^3/\phi^2$ and will ensure that WI takes place in weak dissipative regime. 

\subsection{Linear Potentail: $V(\phi)=V_0+M_0^3\phi$}

\begin{center}
\begin{figure}[!htb]
\includegraphics[width=8.5cm]{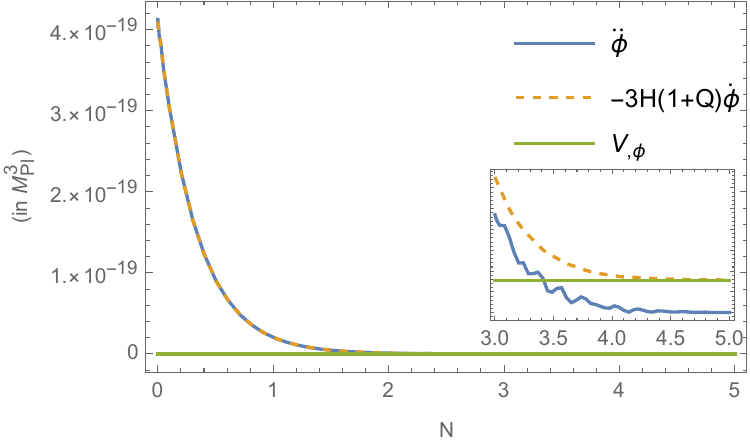}
\caption{The figure depicts the numerical evolution of the acceleration term ($\ddot\phi$), the friction term ($3H(1+Q)\dot\phi$) and the slope term ($V,_\phi$) present in the equation of motion of the inflaton field through an Ultra slow-roll phase in the case of linear potentail. We have chosen the parameters as follows: $V_0=(10^{-4}\,M_{\rm Pl})^4$, $M_0=2.5\times10^{-8}\, M_{\rm Pl}$, $C_\Upsilon=10$ and $g_*=106.75$.}
\label{fig1}
\end{figure}
\end{center}

This potential has been considered in \cite{Dimopoulos:2017ged} to analyze the dynamics of Ultra slow-roll in CI. In this potential, when $V_0\gg M_0^3\phi$, the potential becomes extremely flat, and the system can enter an Ultra slow-roll phase. We chose the parameters $V_0$ and $M_0$ accordingly, which we have quoted in the caption of Fig.~(\ref{fig1}). We see from Fig.~(\ref{fig1}) that initially the acceleration term ($\ddot\phi$) and the friction term ($3H(1+Q)\dot\phi$) dominate over the slope term $(V,_\phi)$ (the graph of $\ddot\phi$ and $3H(1+Q)\dot\phi$ overlaps for the first few e-folds), indicating an Ultra slow-roll phase. Around 3.5 e-folds the acceleration term becomes subdominant, and the dynamics is governed by the friction and the slope terms, indicating an onset of the usual slow-roll phase. 

\begin{center}
\begin{figure}[!htb]
\includegraphics[width=8.5cm]{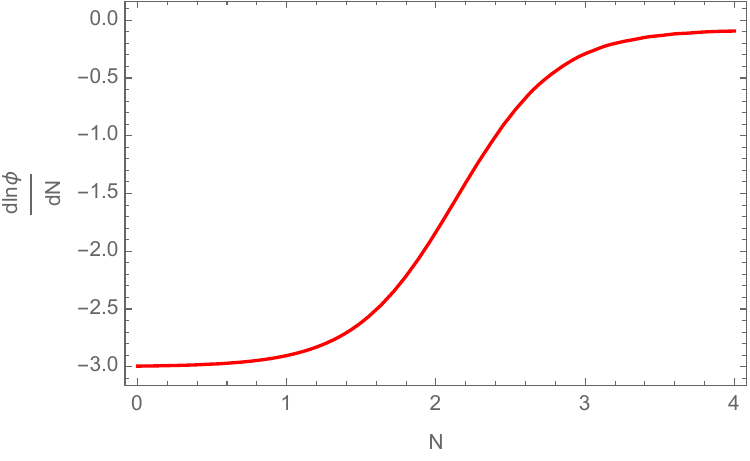}
\caption{Evolution of the thermalization condition given in Eq.~(\ref{thermal-condition}) during Ultra slow-roll in the case of linear potential.}
\label{fig2}
\end{figure}
\end{center}

\begin{center}
\begin{figure}[!htb]
\includegraphics[width=8.5cm]{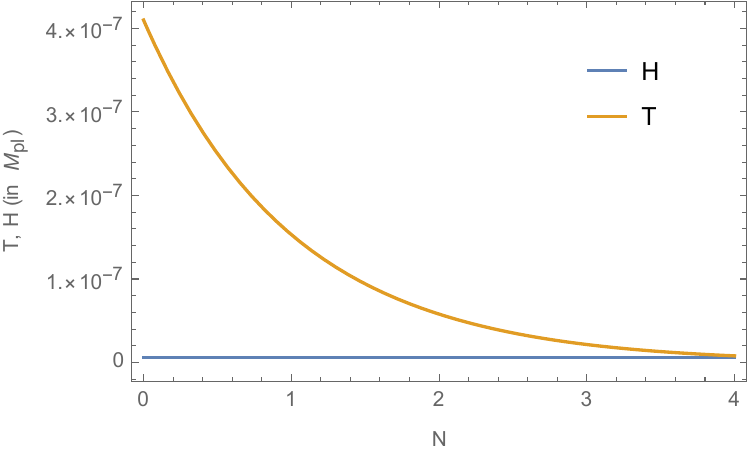}
\caption{Evolution of the temperature and the Hubble parameter during Ultra slow-roll in the case of linear potential.}
\label{fig3}
\end{figure}
\end{center}

 In Fig.~(\ref{fig2}), we note that the thermalization condition during Ultra slow-roll given in Eq.~(\ref{thermal-condition}) is maintained for about the first 1.5 e-folds. However, as the condition for thermalization  differs from the equation of the inflaton field during Ultra slow-roll (Eq.~(\ref{KG-approx})), the thermalization condition cannot be maintained for a long period and the system will veer off from thermal equilibrium, as has been pointed out in the previous section. However, the overall thermalization condition of WI, $T>H$, will remain maintained throughout the Ultra slow-roll phase, as has been depicted in Fig.~(\ref{fig3}). 

\begin{center}
\begin{figure}[!htb]
\includegraphics[width=8.5cm]{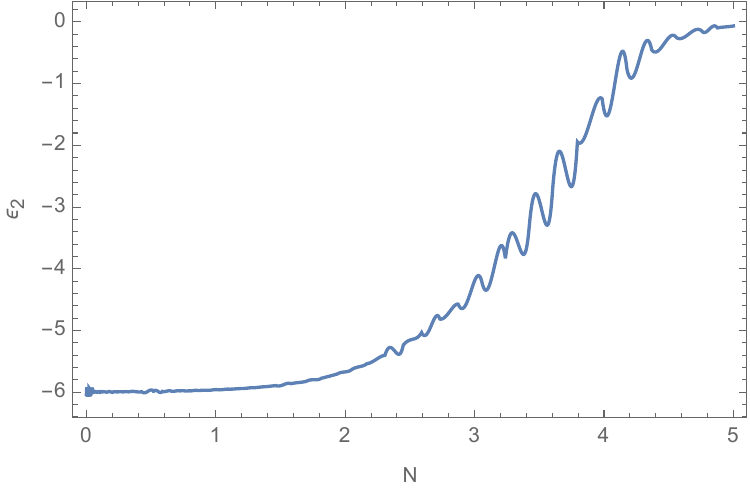}
\caption{Evolution of the second Hubble slow-roll parameter, $\epsilon_2$, during Ultra slow-roll in the case of linear potential.}
\label{fig4}
\end{figure}
\end{center}

\begin{center}
\begin{figure}[!htb]
\includegraphics[width=8.5cm]{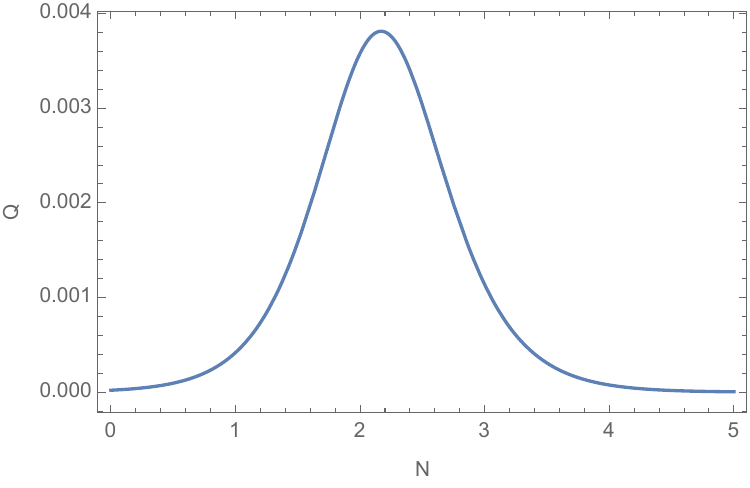}
\caption{Evolution of $Q$ during Ultra slow-roll in the case of linear potentail.}
\label{fig5}
\end{figure}
\end{center}

Fig.~(\ref{fig4}) depicts the evolution of the second Hubble parameter, $\epsilon_2$, during this Ultra slow-roll evolution. We note that the Ultra slow-roll condition, $\epsilon_2<-1$, is maintained till about 4 e-foldings and after that $|\epsilon_2|$ becomes smaller than unity, indicating onset of slow-roll phase. 
Also, Fig.~(\ref{fig5}), where the evolution of $Q$ is shown during Ultra slow-roll, ensures that WI takes place in weak dissipative regime during this Ultra slow-roll phase. 

Hence, we can see that in such a setup an Ultra slow-roll phase can be embedded in a weak dissipative WI model while maintaining thermal equilibrium.

\subsection{Cubic Potential: $V(\phi)=V_0+\left[1+\left(\frac{\phi}{\phi_0}\right)^3\right]$}

The cubic potential was considered in \cite{Pattison:2018bct} while discussing the attractor behaviour of Ultra slow-roll in CI. This potential has an inflection point around $\phi=0$. In WI, if we start near this inflection point region then the system can undergo an Ultra slow-roll phase. It is to note that we cannot analyze the system at the inflection point ($\phi=0$), as the dissipative coefficient ($\Upsilon\propto T^3/\phi^2$) will be ill-defined at this point. We can only analyze the system near about the inflection point. 

\begin{center}
\begin{figure}[!htb]
\includegraphics[width=8.5cm]{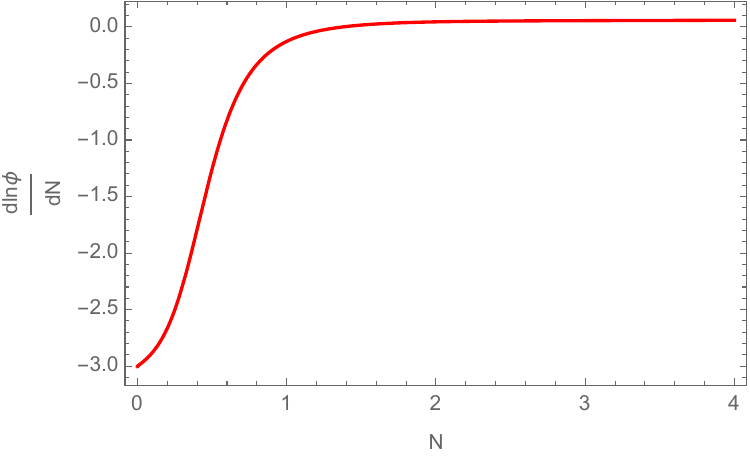}
\caption{Evolution of the thermalization condition given in Eq.~(\ref{thermal-condition}) during Ultra slow-roll in the case of cubic potential.}
\label{fig6}
\end{figure}
\end{center}

\begin{center}
\begin{figure}[!htb]
\includegraphics[width=8.5cm]{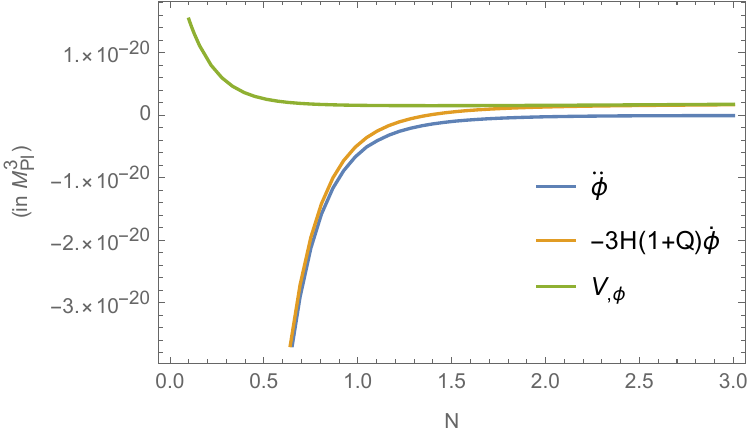}
\caption{The figure depicts the numerical evolution of the acceleration term ($\ddot\phi$), the friction term ($3H(1+Q)\dot\phi$) and the slope term ($V,_\phi$) present in the equation of motion of the inflaton field through an Ultra slow-roll phase in the case of cubic potential. We have chosen the parameters as follows: $V_0=(10^{-4}\,M_{\rm Pl})^4$, $\phi_0=2.5\times10^{-1}\, M_{\rm Pl}$, $C_\Upsilon=10^4$ and $g_*=106.75$.}
\label{fig7}
\end{figure}
\end{center}

\begin{center}
\begin{figure}[!htb]
\includegraphics[width=8.5cm]{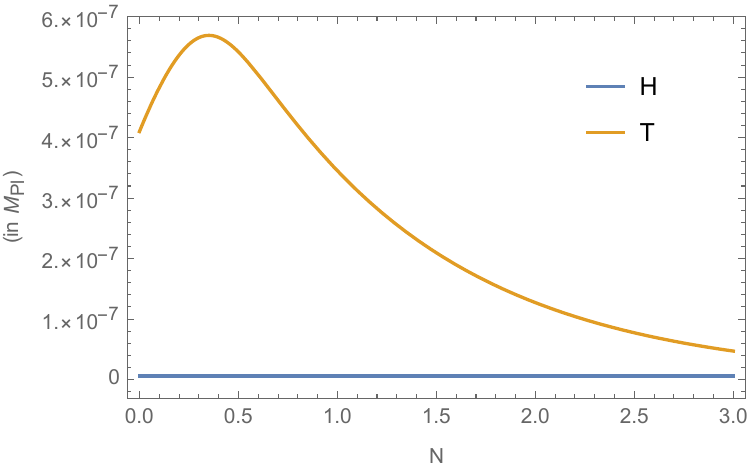}
\caption{Evolution of the temperature and the Hubble parameter during Ultra slow-roll in the case of cubic potential.}
\label{fig8}
\end{figure}
\end{center}

We first note in Fig.~(\ref{fig6}), that in this case the system quickly deviates from the thermalization condition given in Eq.~(\ref{thermal-condition}) as soon as Ultra slow-roll begins. This indicates that the system should not linger in an Ultra slow-roll phase for a long time as that will lead to disruption of thermal equilibrium of the system. In Fig.~(\ref{fig7}), we notice that the coupled equations of the system do not let the Ultra slow-roll phase to last for long, and within nearly 1.5 e-folds the system tends to enter a slow-roll phase. We also note in Fig.~(\ref{fig8}) that the overall thermalization condition of WI ($T>H$) is maintained throughout the Ultra slow-roll phase.

\begin{center}
\begin{figure}[!htb]
\includegraphics[width=8.5cm]{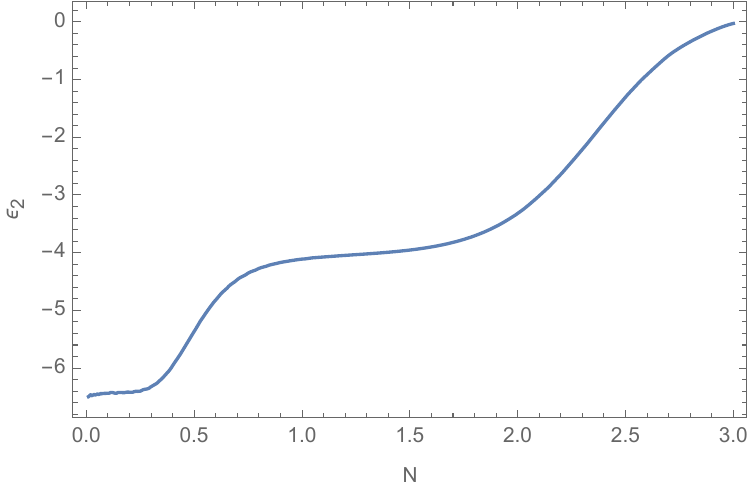}
\caption{Evolution of the second Hubble slow-roll parameter, $\epsilon_2$, during Ultra slow-roll in the case of cubic potential.}
\label{fig9}
\end{figure}
\end{center}

\begin{center}
\begin{figure}[!htb]
\includegraphics[width=8.5cm]{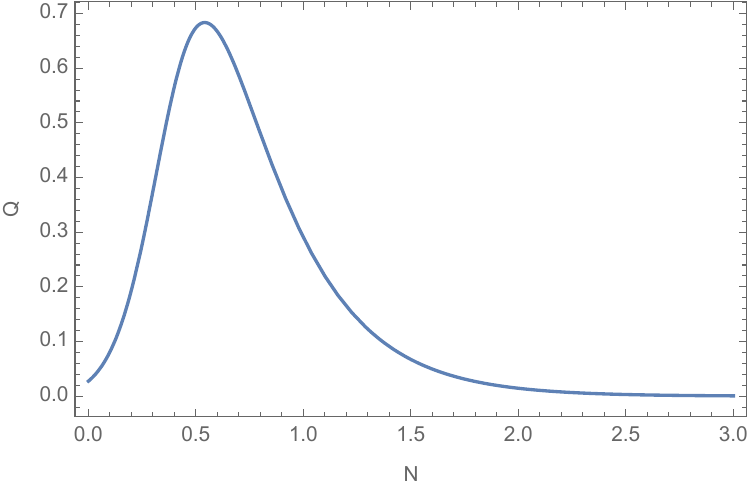}
\caption{Evolution of $Q$ during Ultra slow-roll in the case of cubic potential.}
\label{fig10}
\end{figure}
\end{center}

In Fig.~(\ref{fig9}) we see that the second Hubble slow-roll, $\epsilon_2$, will remain larger than -1 for about 2.5 e-foldings and then we achieve $|\epsilon_2|<1$ indicating an onset of an usual slow-roll phase. Also, Fig.~(\ref{fig10}) ensures that the whole dynamics takes place in a weak dissipative regime. 

Therefore, in this system, too, an Ultra slow-roll phase, though much shorter than the previous scenario, can be realized near an inflection point within a weak dissipative WI model while maintaining the overall thermal equilibrium of the system. 


\section{Discussion and Conclusion}
\label{conclusion}

CI undergoes an Ultra slow-roll phase when the potential becomes extremely flat \cite{{Dimopoulos:2017ged}}. In this manuscript we ask the question what happens to WI, a variant inflationary scenario, in a similar situation. As has been pointed out in the Introduction, a constant, subdominant, nearly thermally equilibrated radiation bath coexists during WI due to the continuous dissipation of energy of the inflaton field into this radiation bath. The coexistence of this radiation bath along with inflaton energy densitiy is the signature of WI which distinguishes it from the standard CI scenario. Therefore, one naturally expects the radiation bath, along with the thermal equilibrium of the system, to be maintained in WI even when the system passes through an extremely flat region of the potential. 

However, we found in this article that WI models with dissipative coefficients solely dependent on the temperature ($\Upsilon\propto T^p$) fail to maintain thermal equilibrium of the system when the system traverses through a very flat region of the potential. It is not yet known what happens to the WI dynamics when the thermal equilibrium is lost. We, therefore, could not further analyze these systems in such circumstances. 

On the other hand, we showed that the overall thermal equilibrium of the system can be maintained throughout a phase when the WI system traverses through an extremely flat region of the potential in cases where the dissipative coefficients are function of both the temperature and the inflaton amplitude. We particularly dealt with the models with dissipative coefficient of the form $\Upsilon\propto T^3/\phi^2$. Such models have been shown to tally well the observations when WI takes place in weak dissipative regime \cite{Benetti:2016jhf, Arya:2017zlb}. Therefore, we treated such models in the weak dissipative regime, and considered two potentials (linear and cubic) with extremely flat regions to demonstrate that an Ultra slow-roll phase can indeed be realised in WI while maintaining the overall thermal equilibrium of the system. 

Though it seems like a positive note to conclude the article, we would like to call attention to our incapability and discomfort to deal with WI systems with dissipative coefficients depending on the temperature alone when such systems encounter an extremely flat region of the potential. We have observed that these systems lose  thermal equilibrium, a signature property of WI. We hope that this article will encourage more research in this field to reveal the true nature of the WI dynamics in such circumstances. We also leave the analysis of the cosmological perturbations during the Ultra slow-roll phase in WI for a future project, which is essential to determine the observational consequences of such a departure from slow-roll during WI. Recently, a numerical code has been developed  \cite{Montefalcone:2023pvh} to study the cosmological perturbations in standard slow-roll WI models. We aim to develop similar codes for an Ultra slow-roll phase in WI in a future project.

\acknowledgements

SD would like to thank Rudnei Ramos for many useful discussions on Warm Inflation from time to time. 



\end{document}